\documentclass[12pt]{article}%
\usepackage{amssymb}
\usepackage{amsmath}
\usepackage{amsthm}
\usepackage{amsfonts}
\usepackage{graphicx}
\usepackage{subfigure}%
\input epsf.tex
\begin{document}
\bigskip\begin{titlepage}
\begin{flushright}
UUITP-07/06\\
hep-th/0605105
\end{flushright}
\vspace{1cm}
\begin{center}
{\Large\bf Stability of flux vacua in the presence of charged black holes\\}
\end{center}
\vspace{3mm}
\begin{center}
{\large
Ulf   H.\   Danielsson{$^1$}, Niklas Johansson{$^2$} and Magdalena Larfors{$^3$}} \\
\vspace{5mm}
Institutionen f\"or Teoretisk Fysik, Box 803, SE-751 08
Uppsala, Sweden \\
\vspace{5mm}
{\tt
{$^1$}ulf.danielsson@teorfys.uu.se\\
{$^2$}niklas.johansson@teorfys.uu.se\\
{$^3$}magdalena.larfors@teorfys.uu.se\\
}
\end{center}
\vspace{5mm}
\begin{center}
{\large \bf Abstract}
\end{center}
\noindent
In this letter we consider a charged black hole in a flux
compactification of type IIB string theory. Both the black hole and
the fluxes will induce potentials for the complex structure moduli.
We choose the compact dimensions to be described locally by a deformed
conifold, creating a large hierarchy. We demonstrate that the presence
of a black hole typically will not change the minimum of the moduli
potential in a substantial way. However, we also point out a couple of
possible loop-holes, which in some cases could lead to interesting
physical consequences such as changes in the hierarchy.\vfill
\begin{flushleft}
May 2006
\end{flushleft}
\end{titlepage}\newpage

\section{Introduction}

Compactifications with non-trivial 3-form fluxes\footnote{For a recent
review with extensive references, see \cite{Grana:2005jc}.} provide an
exciting new way to construct phenomenologically interesting stringy
models of particle physics and cosmology. These models come to terms
with the difficult issue of moduli stabilization
\cite{Giddings:2001yu,Kachru:2003aw} and also provide a possible
explanation for the hierarchy problem of particle physics
\cite{Giddings:2001yu,DeWolfe:2002nn,Klebanov:2000hb}. In
addition there are important consequences for cosmology. The flux
compactifications have lead to a new understanding of the problem of
the cosmological constant \cite{Bousso:2000xa} and can, in addition,
incorporate the process of inflation \cite{Kachru:2003sx}.

The four-dimensional effective theory of a flux compactification depends
heavily on the value to which the complex structure moduli are fixed. For
instance, in the models explaining the hierarchy, the moduli are fixed close
to a conifold point. The distance to the conifold singularity then sets the
hierarchy \cite{Giddings:2001yu}.

As made explicit in \cite{Kallosh:2005ax, Kallosh:2005bj}, the way complex
structure moduli are fixed by fluxes is very analogous to the attractor
mechanism \cite{Ferrara:1995ih,Strominger:1996kf, Ferrara:1996dd} in black
hole physics. This raises the important question whether the presence of a
charged black hole in a flux background can affect the minimum of the
potential for the moduli, and thus affect the hierarchy or other physical
properties of the compactification. In this paper we consider such a
situation. We choose to study fluxes that fix the moduli close to a conifold
point, as in the model explaining hierarchies. In general, adding a black hole
breaks all supersymmetries. Nevertheless, we are able to draw some general conclusions.

At string tree-level, the input from the internal geometry to the moduli
stabilization physics is, in both cases, governed by the prepotential of the
Calabi--Yau. Close to a conifold point, this means that not only the black
hole, but also the flux compactification can be described by a matrix
model\footnote{For the black hole case, this correspondence holds to all
orders in the string loop expansion as described in \cite{Danielsson:2004ti,
Danielsson:2005uz}.}. We will make use of this fact to estimate the shift in
the minimum of the potential caused by the black hole.

Our result is that for a generic choice of 3-fluxes the black hole has very
small impact on the minimum of the potential. The main reason for this is that
the terms in the potential coming from the black hole is suppressed by
$\sim1/(Q^{2}q^{2})$, were $Q$ is integer flux quanta and $q$ is integer black
hole D3-brane charge.

We find a few cases when the above argument might be questioned. These include
situations with fine-tuned flux quanta, and black holes at the end of their
Hawking evaporation, provided the flux quanta and black hole charges are small.

The outline of the paper is as follows. We begin, in section 2, by recalling
the relation between the matrix model free energy and the prepotential of a
conifold limit of a compact Calabi--Yau. With this prepotential all the
interesting attractor phenomena can be studied. Section 3 recalls the relevant
material from black hole attractor physics and flux compactifications, always
keeping our explicit example in mind. In both cases we find effective
four-dimensional potentials for the complex moduli. In section 4 we study the
combined system and compare the relative importance of the potentials. The
paper ends with the conclusions.

\section{The prepotential from the matrix model}

Below we review the established connection [13,14] between the matrix
model and black holes. We do this in order to make the reader think in
matrix model terms when we later discuss both black holes and flux
compactifications. This unified view on the two systems is fruitful
from a conceptual point of view, and will hopefully deepen our
understanding of the physics of flux compactifications. Also, we
explain in more detail the origin of the non-universal terms, crucial
for modelling the embedding of the conifold into a compact Calabi-Yau.

According to \cite{Ooguri:2004zv} there is a direct relation between
BPS black holes in 4D type IIB supergravity and topological strings
propagating on the Calabi--Yau on which the type IIB ten-dimensional
theory is compactified. In \cite{Danielsson:2004ti, Danielsson:2005uz}
this fact was combined with the results of \cite{Ghoshal:1995wm} to
set up a detailed match between the free energy of the $c=1$ matrix
model\footnote{For a nice review of the matrix model, see
\cite{Klebanov:1991qa}.} and the entropy of these extremal black
holes.

We are interested in internal manifolds which have complex moduli such
that they locally look like deformed conifolds. Not only is this the
limit where the matrix model tools are applicable, but it is also the
limit used to explain hierarchies in flux compactifications.

Let us review the calculation in \cite{Danielsson:2005uz} -- with some
more details -- of the free energy in the matrix model paying
attention to large non-universal terms. These non-universal terms give
the main contribution to the entropy, while some of the dependence on
the complex moduli is captured by the universal terms. For
concreteness we start out with a regulated double well potential given
by%
\begin{equation}
V\left(  \lambda\right)  =-\frac{\lambda^{2}}{\alpha^{\prime}}+A\lambda^{4}.
\end{equation}
Using $N$ fermions we fill up the Fermi sea to a level $\mu$ as measured from
the top of the potential. The conifold physics is then described by what is
going on near the top of the potential, while the regulating quartic piece
rounds off the conifold and makes it part of a Calabi--Yau manifold with finite
volume \cite{Danielsson:2005uz}. The details of the regularisation capture
the shape of the manifold away from the conifold tip. Our task is then to find
an expression for the canonical free energy $F_{MM}\left(  N,\beta\right)  $
for the system and its Legendre transform $\mathcal{F}_{MM}\left(  \mu
,\beta\right) $. To accomplish this we express the free energy and the number
of fermions as
\begin{equation}
F_{MM}\left(  N,\beta\right)  =\int^{-\mu}d\varepsilon\varepsilon\rho\left(
\varepsilon\right)  ,
\end{equation}
where $\mu$ is to be substituted for $N$ according to
\begin{equation}
N=\int^{-\mu}d\varepsilon\rho\left(  \varepsilon\right)  ,
\end{equation}
and, where the integration in energy goes from the bottom of the
potential up to the Fermi surface. The density of states is given by%
\begin{equation}
\rho\left(  \varepsilon\right)  =\beta\int_{\lambda_{-}}^{\lambda_{+}}%
\frac{d\lambda}{\sqrt{2\left(  \varepsilon+\frac{\lambda^{2}}{\alpha^{\prime}%
}-A\lambda^{4}\right)  }},
\end{equation}
where the integration limits are the shores of the Fermi sea. It is now a
simple exercise to compute the free energy and we arrive at%
\begin{equation}
\beta\mathcal{F}_{MM}\left(  \mu,\beta\right)  =\frac{1}{\sqrt{\alpha^{\prime
}}}N_{0}^{2}-N_{0}\mu\beta-\sqrt{\alpha^{\prime}}\left(  \beta\mu\right)
^{2}\ln\left(  \mu/\Lambda\right)  , \label{Eq_FreeEnergy}%
\end{equation}
where
\begin{equation}
\Lambda\sim\frac{1}{A\alpha^{\prime2}}%
\end{equation}
is an effective cutoff introduced by the quartic piece of the potential.
$N_{0}$ is the number of fermions needed if we fill the potential all the way
up and is given, through Bohr--Sommerfeldt quantization by%
\begin{equation}
N_{0}\sim\frac{\beta}{A\alpha^{\prime3/2}}.
\end{equation}
Let us explain in some more detail the origin of the various
terms. The last term in expression (\ref{Eq_FreeEnergy}) is well known
and is simply the standard non-analytic universal contribution to the
free energy of the matrix model. In contrast, the first two terms have
an analytic dependence on $\mu$ and do not play any role in the usual
matrix model analysis. Here, however, they are of crucial
importance. The second term is a consequence of the relation%
\begin{equation}
N_{0}=-\left.  \frac{\partial\mathcal{F}_{MM}\left(  \mu,\beta\right)
}{\partial\mu}\right\vert _{\mu=0},
\end{equation}
while the first is obtained from%
\begin{equation}
\mathcal{F}_{MM}\left(  0,\beta\right)  =-F_{MM}\left(  N_{0},\beta\right)
=-\int^{0}d\varepsilon\varepsilon\rho\left(  \varepsilon\right)  \sim-\frac
{1}{\sqrt{\alpha^{\prime}}\beta}N_{0}^{2}.
\end{equation}
Here we have used $\rho\left(  \varepsilon\right)  \sim\beta\sqrt
{\alpha^{\prime}}$ to estimate the bulk density of states. Expressing $N_{0}$
in in terms of the parameters of the problem we finally get%
\begin{equation}
\beta\mathcal{F}_{MM}\left(  \mu,\beta\right)  =\frac{1}{A^{2}\alpha
^{\prime7/2}}\beta^{2}-\frac{\mu\beta}{A\alpha^{\prime3/2}}\beta-\sqrt
{\alpha^{\prime}}\left(  \beta\mu\right)  ^{2}\ln\left(  \mu/\Lambda\right)  .
\end{equation}
The calculation is performed at zero temperature, but as argued in
\cite{Danielsson:2004ti, Danielsson:2005uz}, the relevant temperature of the
matrix model should actually be a multiple of the self-dual temperature in
order to describe the conifold.\footnote{Note that the temperature of the
black hole in space time is still zero, as explained in
\cite{Danielsson:2004ti, Danielsson:2005uz}.} It can be shown, however, that
the general form of the free energy does not change.

Written in the way above, the free energy of the matrix model provides
interesting information about the entropy of four-dimensional black holes. The
canonical free energy $F_{MM}\left(  N,\beta\right)  $ is directly
proportional to the black hole entropy with the various parameters being
related to two sets of electric and magnetic charges. The number of fermions
can be associated with an electric charge $q_{1}\sim N$. The main contribution
to the entropy is given by the analytic piece and is of the form $S\sim N^{2}%
$, while the universal non-analytic piece tells us how the entropy varies
close to the conifold value $N_{0}$. The black hole also have a magnetic
charge given by $p^{0}\sim\beta$. As argued in \cite{Danielsson:2004ti,
Danielsson:2005uz} we can also turn on another magnetic charge, $p^{1}$, which
from the matrix model point of view corresponds to deforming the potential by
a $1/\lambda^{2}$ piece.

Furthermore, as discussed in \cite{Danielsson:2004ti, Danielsson:2005uz}, the
free energy of the matrix model is directly related to the imaginary part of
the prepotential of the four-dimensional supergravity theory. In the next
section we will write down the corresponding prepotential and review how the
attractor equations obtained from the four-dimensional analysis reproduce
known properties of the matrix model and the corresponding black hole. We will
also use the same prepotential to accomplish moduli stabilization through a
flux compactification. In this way we obtain a mapping between quantities of
the matrix model and space time not only in the case of a black hole, but also
for flux compactifications.

\bigskip

\section{Moduli stabilization in type IIB supergravity}

\bigskip

Consider ten-dimensional type IIB supergravity. Neglecting the Chern--Simons
term, the bosonic action is given by\footnote{We use the notations of
\cite{Giddings:2001yu}.}
\begin{align}
S_{IIB}  & = \frac{1}{2\kappa_{10}^{2}}\int d^{10}x\sqrt{-g}\left(  {\cal R} -
\frac{\partial_{M} \tau\partial^{M} \tau}{2(\mbox{Im} \tau)^{2}}-
-\frac{|G_{3}|^{2}}{12\mbox{Im} \tau} - \frac{|\tilde{F}_{5}|^{2}}{4\cdot5!}
\right) .
\end{align}
Here $G_{3}=F_{3}-\tau H_{3}$ and $\tau=\frac{i}{g_{s}}+C_{0}$ is the axio-dilaton.

We will study compactifications of this theory to four dimensions, letting the
internal dimensions be a (possibly conformal) Calabi--Yau manifold $Y$.
Specifically we will assume a complex structure moduli space $\mathcal{M}$ of
complex dimension one, and that we are close to a conifold point. Let
$\{A^{I}, B_{I}\}$ $I = 0,1$ be a symplectic basis of $H_{3}(Y)$, so that
$A^{1}$ is the conifold cycle. Furthermore, let $\{\alpha_{I}, \beta_{I}\}$ be
a basis of $H^{3}(Y)$ so that, as usual,
\begin{align}
\oint_{Y} \alpha_{J} \wedge\beta^{I} = \oint_{A^{I}} \alpha_{J} = \oint_{B_{J}%
}\beta^{I} = \delta^{I}_{J}.
\end{align}
The periods of the holomorphic 3-form $\Omega$ are defined by
\begin{align}
X^{I}  & = \oint_{A^{I}} \Omega\\
F_{J}(X^{I})  & = \oint_{B_{J}}\Omega.
\end{align}
Let us work in a K\"{a}hler gauge in which $X^{0} = V^{1/2}$ and $X^{1} =
V^{1/2}z$. $V$ is the (unwarped) volume of the Calabi--Yau, and $z$ is the
coordinate on $\mathcal{M}$ vanishing at the conifold point.

Close to the conifold the prepotential is given by
\begin{equation}
\label{prepotential}F=ia_{1}(X^{0})^{2}+a_{2}X^{0}X^{1}+ia_{3} \left(
X^{1}\right) ^{2}\ln\frac{X^{1}}{X^{0}},
\end{equation}
where other terms of order $\mathcal{O}(z^{2})$ have been neglected,
and the $a_{i}$ are numerical coefficients depending on the
Calabi--Yau geometry.  Specifically, $a_{3} = -1/4\pi$. Note that this
prepotential has exactly the same functional form as the matrix model
free energy.

Let us now study in turn how wrapped branes and fluxes behave on this geometry.

\subsection{A black hole attractor}

The presence of a black hole consisting of wrapped D3-branes generates an
effective potential for the complex structure moduli. The potential is induced
by the 5-form field strength $\tilde{F}_{5}$ sourced by the black hole
charges. The metric is an unwarped product between a four-dimensional part and
a Calabi--Yau part whose complex structure depends on the space-time point.

We write the four-dimensional part of the black hole metric in the form%
\begin{equation}
\tilde{g}_{\mu\nu}^{(4)}dx^{\mu}dx^{\nu}=-e^{2u\left(  \sigma\right)  }%
dt^{2}+\frac{e^{-2u\left(  \sigma\right)  }c^{4}d\sigma^{2}}{\sinh^{4}\left(
c\sigma\right)  }+\frac{e^{-2u\left(  \sigma\right)  }c^{2}}{\sinh^{2}\left(
c\sigma\right)  }d\Omega^{2},
\end{equation}
where $c\rightarrow0$ is the extremal limit. Here $\sigma$ goes from $-\infty$
(horizon) to $0$ (spatial infinity). Furthermore $u \sim c\sigma$ as
$\sigma\rightarrow-\infty$.

In the notation of \cite{Denef:2000nb} the field strength is given by
\begin{equation}
\tilde{F}_{5} = \mathcal{F+}\ast\mathcal{F=}\sin\theta d\theta\wedge
d\phi\wedge\Gamma+ e^{2u}dt\wedge d\sigma\wedge\hat{\Gamma}.
\end{equation}
Here $\Gamma$ is a 3-form corresponding to the black hole charge, and
$\hat{\Gamma} = *_{6} \Gamma$ is its six-dimensional Hodge dual. In
particular, if the D3-brane wraps the cycles $A^{I} (B_{I})$ $q_{I} (p^{I})$
times, then $\Gamma= \pi(\alpha^{\prime})^{2}(p^{I}\alpha_{I} + q_{I}
\beta^{I})$ and, consequently,
\begin{align}
\oint_{S^{2} \times A^{I}} \tilde{F}_{5} = ((2\pi)^2\alpha^{\prime})^{2} p^{I} &
\mbox{ and }\oint_{S^{2} \times B_{I}} \tilde{F}_{5} = ((2\pi)^2\alpha^{\prime
})^{2} q_{I},
\end{align}
for any space-like $S^{2}$ enclosing the location of the D3-branes.

In the four-dimensional effective action, this field strength gives rise to
the term
\begin{align}
S_{pot}  & = -\frac{1}{2\kappa_{10}^{2}}\int d^{10}x \sqrt{-g} \frac
{|\tilde{F}_{5}|^{2}}{4\cdot5!} = -\frac{1}{2\kappa_{10}^{2}} \int
d\mbox{Vol}_{4} \frac{1}{r^{4}} \oint_{Y}\Gamma\wedge\hat{\Gamma}=\\
& =-\frac{1}{2\kappa_{10}^{2}}\int d\mbox{Vol}_{4} \frac{V_{bh}\left(
z\right) }{r^{4}},\nonumber
\end{align}
where we reinserted the usual radial coordinate $r$. In the case of a BPS
black hole, the potential $V_{bh}$ can be obtained via the Gukov--Vafa--Witten
(GVW) superpotential $W_{bh} = \int_{Y} \Omega \wedge \Gamma$
\cite{Gukov:1999ya} as the usual ${\cal N}=2$ scalar potential
\begin{equation}
V_{bh}\left(  z\right)  =e^{K_{bh}}\left(  \mathcal{G}_{bh}^{z\bar{z}}
D_{z}W_{bh} D_{\bar{z}}\overline{W}_{bh}+ \left\vert W_{bh}\right\vert ^{2}\right)
,
\end{equation}
where $\mathcal{G}_{bh}$ is the metric derived from the K\"{ahler} potential
$K_{bh} = -\ln i\left(  \bar{X}^{I}F_{I}-X^{I}\bar{F}_{I}\right) $ on
$\mathcal{M}$. Using equation (\ref{prepotential}), it is straightforward to
express $W_{bh}$ and $K_{bh}$ in terms of the geometrical coefficients $a_{i}$
and the black hole charges. Explicitly, with our gauge choice, we
obtain\footnote{For notational simplicity we ignore factors of $\pi$.}
\begin{equation}
W_{bh}\left(  z\right)  = V^{1/2}(\alpha^{\prime})^{2} w(z)
\end{equation}
with
\begin{equation}
w= q_{0}-2ia_{1}p^{0}-a_{2}p^{1}+\left(  q_{1}-a_{2}p^{0}-ia_{3}p^{1}\right)
z+ia_{3}p^{0}z^{2}-2ia_{3}p^{1}z\ln z .
\end{equation}
Furthermore, we have the K\"{a}hler potential $K_{bh}=K_{bh}\left(  z,\bar
{z}\right) $ given by%
\begin{equation}
e^{-K_{bh}}= Vk(z,\bar{z}) = V \left(  4a_{1}-a_{3}\left(  z-\bar{z}\right)
^{2}+2 a_{3} \left\vert z\right\vert ^{2}\ln\left\vert z\right\vert ^{2}
\right) .
\end{equation}
It is now easy to see that the usual matrix model results are reproduced. If
we minimize the potential through $\partial_{z}V_{bh}=0,$ we find%
\begin{equation}
D_{z}W_{bh}= V^{1/2} \frac{kw_{z}-k_{z}w}{k}=0.
\end{equation}
This is just the attractor equations for the complex structure moduli.
Focusing on the black hole corresponding to the undeformed matrix model (it is
easy to generalize to the general case), we have $q_{0}=p^{1}=0$ and the
attractor equations tell us that, to first order in $z$,%
\begin{equation}
q_{1}=a_{2}p^{0}-2a_{3}p^{0}x\ln x-a_{3}p^{0}x,
\end{equation}
where $z=ix$ and $x$ is real. Note that $z=\frac{X^{1}}{X^{0}} \sim i\mu$.
This is nothing else than the formula for the number of fermions $q_{1}\sim N$
near its critical value given by $p^{0}\sim\beta$, when we fill up the Fermi
sea towards the top of the potential.

\bigskip

\subsection{Flux compactifications and hierarchies}

\bigskip

Warped geometries have played a crucial role in the construction of
realistic phenomenological models. The reasons are twofold. On the one
hand the introduction of 3-form fluxes, $F_{3}$ and $H_{3}$, on the
compact manifold works just like the introduction of space time
filling D3-branes. These branes appear as point sources on the compact
manifold and correspond to deep throats of warped geometry. The
warping introduces a relative redshift between various points on the
compact manifold which can be used to explain hierarchies of scales.

On the other hand, the fluxes introduce potentials for the complex moduli of
the Calabi--Yau manifold. This happens quite analogously to the black hole
case. One difference, however, is that the potential now receives
contributions not only from the 3-form flux term in the action, but also from
the 5-form and the Einstein--Hilbert term. Through the equations of motion,
these terms can be rewritten in terms of the fluxes.

The potential part of the effective action becomes \cite{DeWolfe:2002nn}
\begin{align}
S_{pot} = -\frac{1}{2\kappa_{10}^{2}}\int d\mbox{Vol}_{4} \oint_{Y}
\frac{e^{4A}}{2 \mbox{Im}\tau}G_{3} \wedge\left(  *_{6} \overline{G}_{3} +
i\overline{G}_3 \right)  = -\frac{1}{2\kappa_{10}^{2}}\int d\mbox{Vol}_{4}
V_{f}\left(  z\right) .
\end{align}
Also in this case, the form of the effective potential for the moduli is
governed by the geometry of the internal manifold. It it given by the usual
${\cal N}=1$ scalar potential%
\begin{equation}
V_{f}\left(  z\right)  =e^{K_{f}}\left(  \mathcal{G}^{A\overline{B}} D_{A}W_{f}
D_{\overline{B}}\overline{W}_{f}- 3\left\vert W_{f}\right\vert ^{2}\right)  .
\end{equation}
where $W_{f} = \int_{Y} \Omega \wedge G_3$ again is the GVW superpotential.
The indices $A,B$ go over $z$, $\tau$ and the volume modulus $\rho$. The
K\"{a}hler potential $K_{f}$ now also depends on the axio-dilaton and on the
volume modulus:
\begin{equation}
K_{f} = -\ln\left[  -i(\tau-\bar{\tau})\right] -3\ln\left[  -i(\rho-\bar{\rho
})\right]  -\ln\left[  -i\oint_{Y} e^{-4A}\Omega\wedge\overline{\Omega} \right] .
\end{equation}
The coefficient of the $\rho$ term shows that the K\"{a}hler potential
is of no-scale form, as noticed in \cite{Giddings:2001yu}. From here
it is a straightforward calculation to obtain the behaviour of $V_{f}$
in terms of the flux quanta and the geometrical parameters $a_{i}$. We
return to this in the next section.

We choose non-zero fluxes such that%
\begin{align}%
%TCIMACRO{\doint _{B_{1}}}%
%BeginExpansion
{\displaystyle\oint_{A^{1}}}
%EndExpansion
F_{3}  &  =(2\pi)^{2} \alpha^{\prime}P^{1}\\%
%TCIMACRO{\doint _{A^{1}}}%
%BeginExpansion
{\displaystyle\oint_{B_{1}}}
%EndExpansion
H_{3}  &  = - (2\pi)^{2} \alpha^{\prime}Q_{1},
\end{align}
where $P^{1}$ and $Q_{1}$ are integers. If we do this we end up with a
superpotential of the same form as in \cite{Giddings:2001yu}. Specifically, we
have (still ignoring factors of $\pi$)%
\begin{equation}
W_{f}\left(  z\right) = V^{1/2} \alpha^{\prime}\left( -a_{2}P^{1}+\left(  \tau
Q_{1}-ia_{3}P^{1}\right)  z-2ia_{3}P^{1}z\ln z \right) .
\end{equation}
The only difference from the analysis of the black hole is to keep track of
the complex coupling $\tau$ that multiplies the $H_{3}$ fluxes. The attractor
equations tell us, in the limit of small $z$, that%
\begin{equation}
\tau Q_{1}-2ia_{3}P^{1}(\ln z + 3/2)\sim0.
\end{equation}
This leads to an exponentially small modulus $z\sim e^{-Q_{1}/g_{s}P^{1}}$.
Actually, we must also turn on an $H_{3}$ flux $P^{0}$ through the $A^{0}$
cycle in order to satisfy the axio-dilaton equation $D_{\tau}W=0$ at minimum.
This will fix the string coupling as explained in \cite{Giddings:2001yu}.

As argued in \cite{Giddings:2001yu} this procedure gives a possible
explanation for a large hierarchy through the relation between the moduli and
the warp factor. The conifold equation is given by
\begin{equation}
y_{1}^{2}+y_{2}^{2}+y_{3}^{2}+y_{4}^{2}=z,
\end{equation}
where a non-zero modulus $z$ cuts off the deep throat. Hence the warp factor
can not become arbitrarily small.

We note the similarity with the black hole case. With the particular charges
we have chosen the black hole modulus became purely imaginary, while it became
real in the flux case (if $\tau=\frac{i}{g_{s}}$). However, an arbitrary $\tau$
yields a complex modulus. Similarly, in the black hole case, a non-zero
$p^{1}$ charge leads to a complex modulus. This would correspond to the
deformed matrix model.

\bigskip

\section{A black hole in a flux background}

\bigskip

We now come to the main topic of our discussion: a combined analysis where we
consider a black hole in a flux background.

There are topological restrictions against introducing D3-branes 
in backgrounds with fluxes \cite{Maldacena:2001xj,Witten:1998cd,
Freed:1999vc,Freed:2000ta}. Most importantly, the 3-fluxes $H_3$ 
and $F_3$ need to be cohomologically trivial on the world-volume 
of the brane. This reduces the space of possible charges of the 
black hole. We will consider completely general charges, and only
implicitly assume that they can be consistently introduced  into 
the background in question.

According to the attractor mechanism, complex structure moduli are drawn to
fixed values on the horizon of an extremal black hole. This is only true,
however, if there is no other contributions to the potential for the complex
structure moduli. For instance, in a flux compactification there is a possible
conflict with the value determined far away from the black hole through the
fluxes. We can expect a competition between the potential as given by the
fluxes and the potential induced by the black hole. The physical question we
would like to address is whether the black hole, in an appreciable way, can
affect where the fluxes lock the moduli.

We imagine a flux compactification where the moduli are fixed at the minimum
of $V_{f}\left(  z\right) $. That is, we fix $z=z_{f}$ such that $\partial_{z}
V_{f}\left(  z_{f}\right)  = 0$. This remains true even if there is a black
hole present provided we are far away from the black hole. What happens if we
move in closer? Eventually the black hole potential $V_{bh}\left(  z\right) $
will start to play a role and we need to consider the combined system.

To exactly solve for a black hole in a flux compactification is certainly a
very complicated task. In principle we should start with an ansatz of the form%
\begin{equation}
ds^{2}=e^{2A\left(  y,\sigma\right)  }\tilde{g}_{\mu\nu}^{(4)}dx^{\mu}dx^{\nu
}+e^{-2A\left(  y,\sigma\right)  }\tilde{g}_{nm}^{(6)}dy^{n}dy^{m},
\end{equation}
where the four-dimensional part is the same as before and we have allowed for
a warp factor depending on space time. We will not go through such an
analysis. What we will do instead is simply to estimate when the two competing
effects are of comparable order and if and when interesting new physics can
occur. To do this we just need the expressions for the respective potentials.
The total effective potential piece, ignoring back reaction on the flux term
from the black hole piece and vice versa, is given by%
\begin{equation}
\label{eq_SW}S_{pot} = -\frac{1}{2\kappa_{10}^{2}}\int d\mbox{Vol}_{4} \left(
\frac{1}{r^{4}} V_{bh}\left(  z\right)  + V_{f}\left(  z\right)  \right) .
\end{equation}
When examining this expression we disregard effects of the warping. Since the
warped throat is small compared to the bulk, the warping ought to cancel out
when integrating over the whole internal manifold \cite{Giddings:2005ff}.

It is clear from (\ref{eq_SW}) that the effect of the black hole is
largest at the horizon. There the black hole potential is suppressed
by a factor of $R^{-4}$, where $R$ is the black hole radius. Thus, for
any macroscopic black hole, it will be substantially suppressed. In
order to study the charge dependence of the suppression we note that
for an extremal black hole, the radius is proportional to the charge
$q$ of the black hole,\footnote{The exact constant of of
proportionality depends on the geometry and size of the internal
dimensions.} $R \sim q$. The potentials themselves are proportional to
the square of the corresponding flux quanta $Q$ and charge. We
therefore expect that the effect of the black hole is suppressed by a
numerical factor $\sim1/(Q^{2}q^{2})$. Thus the effect on the minimum
of the potential should be negligible\footnote{This analysis might not
apply to charged black holes that classically have vanishing horizon
area. When higher derivative terms in the action are taken into
account, these black holes acquire a string scale horizon area that
scales as $\sim q$ \cite{Sen:1995in,Dabholkar:2004yr}.}.

This qualitative argument might however go wrong if the functional forms of the
two potentials $V_{f}$ and $V_{bh}$ differ substantially. That this could be
the case can be seen from
\begin{equation}
V_{f}\left(  z\right)  =e^{K_{f}}\left(  \mathcal{G}_{f}^{z\bar{z}} D_{z}W_{f}
D_{\bar{z}}\overline{W}_{f}+ \mathcal{G}_{f}^{\tau\bar{\tau}} D_{\tau}W_{f}
D_{\bar{\tau}}\overline{W}_{f} \right) ,
\end{equation}
\begin{equation}
V_{bh}\left(  z\right)  =e^{K_{bh}}\left(  \mathcal{G}_{bh}^{z\bar{z}}
D_{z}W_{bh} D_{\bar{z}}\overline{W}_{bh}+\left\vert W_{bh}\right\vert ^{2}\right) ,
\end{equation}
where we used the no-scale behaviour to eliminate $-3|W_{f}|^{2}$. A simple
calculation shows that $\mathcal{G}_{f}^{\tau\bar{\tau}} D_{\tau}W_{f}
D_{\bar{\tau}}\overline{W}_{f}\sim|W_{f}(\tau\rightarrow\bar{\tau})|^{2}$. Thus,
almost identical terms appear in both potentials. The only thing that might be
a concern is if the dominant term in $\partial_{z} V_{f}$ vanishes.

Let us therefore study these expressions more closely, using the explicit
prepotential (\ref{prepotential}). For both the flux and the black hole case
let
\begin{align}
W \sim w(z)=A_{0}+A_{1}z+A_{2}z^{2}+A_{3}z\ln z\\
e^{-K} \sim k(z)=B_{0}+B_{1}\left(  z-\bar{z}\right)  ^{2}+B_{2}z\bar{z}\ln
z\bar{z}%
\end{align}
where the $A_{i}$ and $B_{i}$ are combinations of flux quanta/charges and
geometrical constants, which are linear in the charges. Since we are
interested in where the modulus is fixed, we study $\partial_{z} V$. Using the
above expressions the leading terms are\footnote{Note that these expressions
are valid for any charge/flux configuration. In particular, $z_{bh}$ need not
lie close to the conifold point.}
\begin{align}
\label{DWDW}\frac{\partial}{\partial z}e^{K}W\overline{W}  &  \sim\frac{A_{0}%
}{B_{0}}\left(  A_{1}+A_{3}(\ln z + 1)\right) \\\label{DWDW2}
\frac{\partial}{\partial z}e^{K}{\cal G}^{z\bar{z}}D_{z}WD_{\bar{z}}\overline{W}  &
\sim\frac{1}{z} \left( 2A_{3} \frac{\left(  A_{1}+A_{3}(\ln z+1)\right)
}{B_{2}\ln z\bar{z}}-\frac{\left(  A_{1}+A_{3}(\ln z+1)\right)  ^{2}}%
{B_{2}\left(  \ln z\bar{z}\right) ^{2}} \right) .
\end{align}
We see that (for small $z$) the dominant contributions to the
derivative of the potentials come from $DW\overline{D}\overline{W}$ in
both cases. Let us now add the two potential contributions and solve
for $z$. Since $\ln z\bar{z}$ is a large number we need only consider
the first term in (\ref{DWDW2}). Thus we wish to solve
\begin{equation}
\label{the}R^{4} A_{3f}\left( A_{1f}+A_{3f}(\ln z+1)\right)  + A_{3bh}\left(
A_{1bh}+A_{3bh}(\ln z+1)\right) =0.
\end{equation}
Now we use that $\left( A_{1f}+A_{3f}(\ln z_f + 1)\right)  = 0$ is the zeroth
order attractor equation. Thus, we have that
\begin{equation}
A_{1f} + A_{3f}(\ln z + 1) = A_{1f} + A_{3f}(\ln z_{f} + \ln(z/z_{f}) + 1)
\sim A_{3f}\ln(z/z_{f}).
\end{equation}
Solving (\ref{the}) now yields
\begin{equation}
\ln z - \ln z_{f} = -\frac{A_{3bh}^{2}}{A_{3f}^{2} R^{4}}\left(  \ln z_{f} +
\frac{1+A_{1bh}}{A_{3bh}}\right) .
\end{equation}
This equation shows that indeed, for the generic case, $z$ will be fixed close
to $z_{f}$. This is because the prefactor of the right-hand side is
$\sim1/(Q^{2} q^{2})$.

However, note that $\ln z_{f}$ is typically rather large: to create a
hierarchy of the weak and Planck energy scales of order $\sim10^{-15}$
we need $\ln z_{f} \sim-100$ \cite{Giddings:2001yu}. Therefore it
might suffice to have $1/(Q^{2} q^{2})$ as small as $1/100$ to change
the fixed value of the modulus by a factor significantly different
from 1. This would correspond to $q \sim1$, and $Q \sim10$. If the
fluxes and charges are small, this could be possible for a black hole
in the end of its evaporation process. For such small charges,
however, the supergravity approximation used in this analysis is
likely to be invalid.

Performing the same analysis for the case $A_{3}=0$, we obtain the leading
contribution to be
\begin{equation}
\partial_{z} V \sim\frac{1}{z}\frac{A_{1}^{2}}{B_{2} (\ln(z\bar{z}))^{2}}.
\end{equation}
Since, from the attractor equations, the constant $A_{1f} = \mathcal{O}%
(z_{f})$ this contribution is generally much smaller than the dominant term
when $A_{3f} \neq0$. Thus if the black hole has a $p^{1}$ charge while there
is no $P^{1}$ flux we have a problem. This is however a fine-tuned case. For a
generic flux compactification such flux will be present.

Our  conclusion is  that  the flux  compactifications generically  are
stable against the introduction  of macroscopic black holes. There are
some special cases where the black hole might be important: notably if
there  is  no $A^{1}$-flux,  or  at  the  end of  Hawking  evaporation
provided the fluxes and charges are  small. In these cases it would be
important to work  out the explicit dependence on the geometry of the
extra dimensions. We will return to this in a future publication.

\bigskip

\section{Conclusions and outlook}

\bigskip

In this paper, we have seen that the complex structure moduli stabilization
provided by type IIB flux compactifications is stable against the introduction
of a charged black hole. We have considered the phenomenologically interesting
case when the type IIB theory is compactified on a local deformed conifold,
and fluxes are chosen such that the complex structure modulus is fixed near
the conifold point. By wrapping D3-branes around cycles of the internal
manifold, we have added a black hole to this picture. The leading terms for
the moduli-fixing potentials show that the black hole effect is negligible in
the generic case. In particular, the black hole contribution is suppressed by
$1/(Q^{2}q^{2})$ where $Q$ is integer flux quanta, and $q$ integer black hole
charges. We find a few exceptional cases where the conclusion might not hold;
for instance if there is no $A^{1}$-flux, or at the end of Hawking evaporation
provided the fluxes and charges are small.

So far, we have analysed how a black hole influences the moduli fixing of a
flux compactification. It would also be interesting to study the reversed
question, i.e. how the black hole behaves in a flux background. We have
already seen that the black hole attractor mechanism is changed by the fluxes,
since these fix the complex structure moduli to a new point in moduli space.
Since, for example, the horizon area depends on the value of the moduli at the
black hole horizon we might expect that the black hole physics is altered.

Furthermore, our analysis has been qualitative, and quantitative results would
be very interesting. To achieve this, the full ten-dimensional equations of
motion need to be solved. One would then be in a position to study the moduli
fixing exactly and, e.g., how the warping depends on space-time.

We have also seen that, via the internal geometry there is a correspondence
between a matrix model and this flux compactification. It would be interesting
to see if a matrix model approach could be applied to other aspects of such
effective theories. In particular it would be very interesting to investigate
whether, as in the black hole case, some matrix model could provide quantum
corrections to the compactified theory. Such a relationship could possibly be
found via a topological string theory on the generalized Calabi--Yau manifold
used in the compactification.

\section*{Acknowledgments}

We thank Marcel Vonk for discussions. The work was supported by the Swedish
Research Council (VR).

\bigskip

\bigskip\noindent{\textbf{Note added:}}

\noindent{While this work was being completed the related paper
\cite{Green:2006nv} appeared. There the authors consider a general class of
potentials without the restrictions due to flux compactifications which we
impose in our work.}

\end{document}